\documentclass[conference]{IEEEtran}
\IEEEoverridecommandlockouts
\usepackage{cite}
\usepackage{tikz}
\usepackage{amsmath,amssymb,amsfonts}
\usepackage{algorithmic}
\usepackage{graphicx}
\usepackage{textcomp}
\usepackage{xcolor}
\usepackage{booktabs}
\usepackage{hyperref}

\def\BibTeX{{\rm B\kern-.05em{\sc i\kern-.025em b}\kern-.08em
    T\kern-.1667em\lower.7ex\hbox{E}\kern-.125emX}}
\begin{document}

\title{Investigating the Effectiveness of Explainability Methods in Parkinson's Detection from Speech
\thanks{\textsuperscript{*}Both authors contributed equally to this research. For these authors, the order is alphabetical
\\ 
\\ This work has been submitted to the IEEE for possible publication. Copyright may be transferred without notice, after which this version may no longer be accessible.}}

\author{
    \IEEEauthorblockN{
        Eleonora Mancini\textsuperscript{* 1}, 
        Francesco Paissan\textsuperscript{* 2,3}, 
        Paolo Torroni\textsuperscript{1}, 
        Mirco Ravanelli\textsuperscript{3,4,6}, 
        Cem Subakan\textsuperscript{3,4,5}
    }
    \IEEEauthorblockA{
        \textsuperscript{1}DISI, University of Bologna, Italy 
        \textsuperscript{2}Fondazione Bruno Kessler, Italy 
        \textsuperscript{3}Mila-Québec AI Institute, Canada  \\
        \textsuperscript{4}Concordia University, Canada 
        \textsuperscript{5}Laval University, Canada
        \textsuperscript{6} Université de Montréal, Canada
    } 
}
\maketitle

\begin{abstract}
Speech impairments in Parkinson's disease (PD) provide significant early indicators for diagnosis. While models for speech-based PD detection have shown strong performance, their interpretability remains underexplored. This study systematically evaluates several explainability methods to identify PD-specific speech features, aiming to support the development of accurate, interpretable models for clinical decision-making in PD diagnosis and monitoring. Our methodology involves (i) obtaining attributions and saliency maps using mainstream interpretability techniques, (ii) quantitatively evaluating the faithfulness of these maps and their combinations obtained via union and intersection through a range of established metrics, and (iii) assessing the information conveyed by the saliency maps for PD detection from an auxiliary classifier. Our results reveal that, while explanations are aligned with the classifier, they often fail to provide valuable information for domain experts.
\end{abstract}


\section{Introduction}
Parkinson's disease (PD) is a progressive neurodegenerative disorder primarily marked by the deterioration of dopaminergic neurons in the midbrain. This degeneration leads to a range of motor and non-motor symptoms, including tremors, bradykinesia, cognitive impairment, and depression \cite{hornykiewicz1998biochemical,poewe2017parkinson,de2000prevalence}. Importantly, during the prodromal stages of PD, patients often start to exhibit speech impairments, which can serve as early indicators of the disease \cite{pinto2004treatments,rusz2013imprecise}. Given the non-invasive, cost-effective, and automated nature of speech analysis, researchers have increasingly focused on this approach as a promising avenue for the early detection of Parkinson's disease \cite{laquatra24_interspeech}.

Despite substantial advances in PD classification using speech analysis, model interpretability remains underexplored. In clinical settings, explainable AI (XAI) is essential for providing clear, clinically relevant insights, crucial for the acceptance of automated systems in clinical trials. Many XAI techniques exist for interpreting model predictions, with \textit{post-hoc} explanation methods among the most widely used \cite{molnar2022}. Here, we focus on two different sets of approaches: Perturbation-based and Gradient-based post-hoc explanation methods. Perturbation-based methods assess feature importance by modifying input data, while gradient-based methods use gradients of predictions with respect to inputs \cite{MERSHA2024128111}. Both approaches support local and global explanations, are model-agnostic, and offer valuable insights into the model’s behaviour. This paper systematically evaluates several key perturbation and gradient-based techniques to determine their effectiveness in highlighting PD-relevant speech features, aiming to enhance transparency and clinical utility in PD detection from speech.

Our experimental results show that, although explanations are aligned with the classifier, they often fail to provide insights that are truly informative for domain experts. These methods may lack the level of interpretability required for practical use, emphasizing the need for more effective explainability approaches that connect model behavior with human understanding in specialized domains.

\section{Related Work}
Various studies have explored automated techniques for identifying speech impairments and using them to predict the progression of Parkinson's disease. For example, researchers have applied Convolutional Neural Networks (CNNs) on spectrograms of patients' speech to detect dysarthria, assess its severity and distinguish between PD patients and healthy controls \cite{sonawane2021speech,rios2022end}. Some approaches use a combination of one-dimensional and two-dimensional CNNs to capture temporal and frequency information \cite{vasquez2021transfer,quan2022end}. 
%Additionally, vocal biomarkers associated with PD have been investigated, although they are not always directly used in feature modeling. 
However, some studies suggest that while these biomarkers are effective, models employing non-interpretable features often outperform those built on more interpretable characteristics \cite{FAVARO2023107559}. 

Unfortunately, much of the existing research prioritizes performance metrics over interpretability. As a result, although interpretable speech-based biomarkers have been shown to be useful for Parkinson's diagnosis \cite{10.3389/fneur.2023.1142642}, little effort has been devoted to comprehensive analysis of explainability in PD detection models. Among the most commonly used XAI methods are perturbation and saliency-based approaches. Some studies employed methods such as GradCAM and EigenCAM to visualize important features. However, these works do not include rigorous quantitative validation of their interpretability. %\cite{s24144625}. 
Other approaches, including those using SHAP \cite{maffia2024automatic}, face challenges in explainability due to the complex input features—such as Mel-frequency cepstral coefficients (MFCCs)—that do not directly correlate with auditory concepts relevant to PD.

\begin{figure*}
    \centering
    \resizebox{0.9\textwidth}{!}{
    \begin{tikzpicture}
        \node (org) {\includegraphics[width=0.26\textwidth]{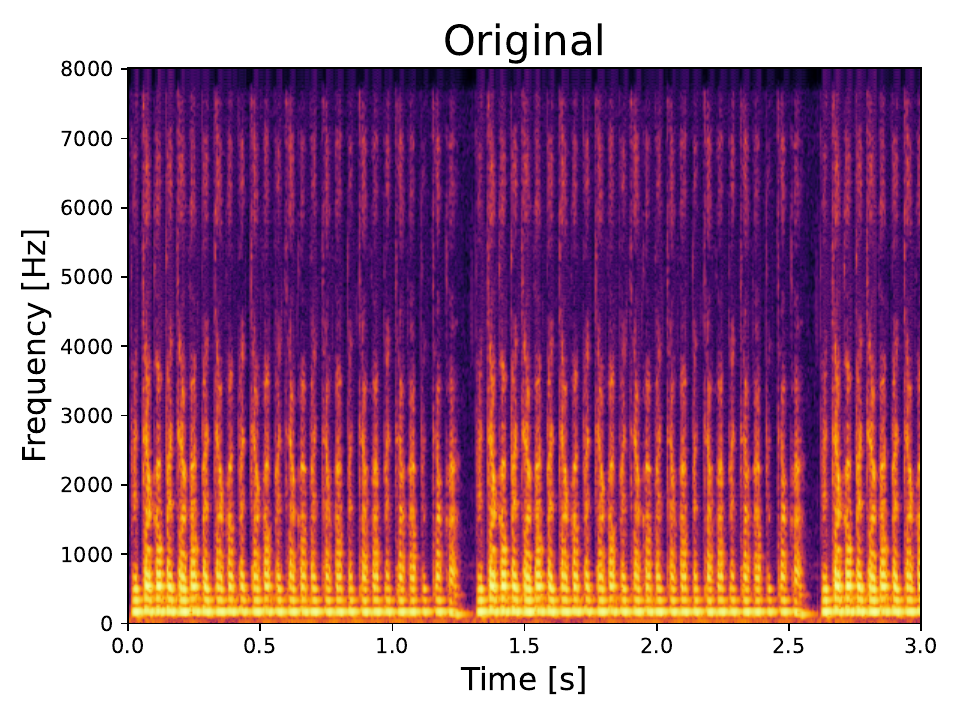}};
        \node[right of=org, xshift=3.5cm] (p1) {\includegraphics[width=0.26\textwidth]{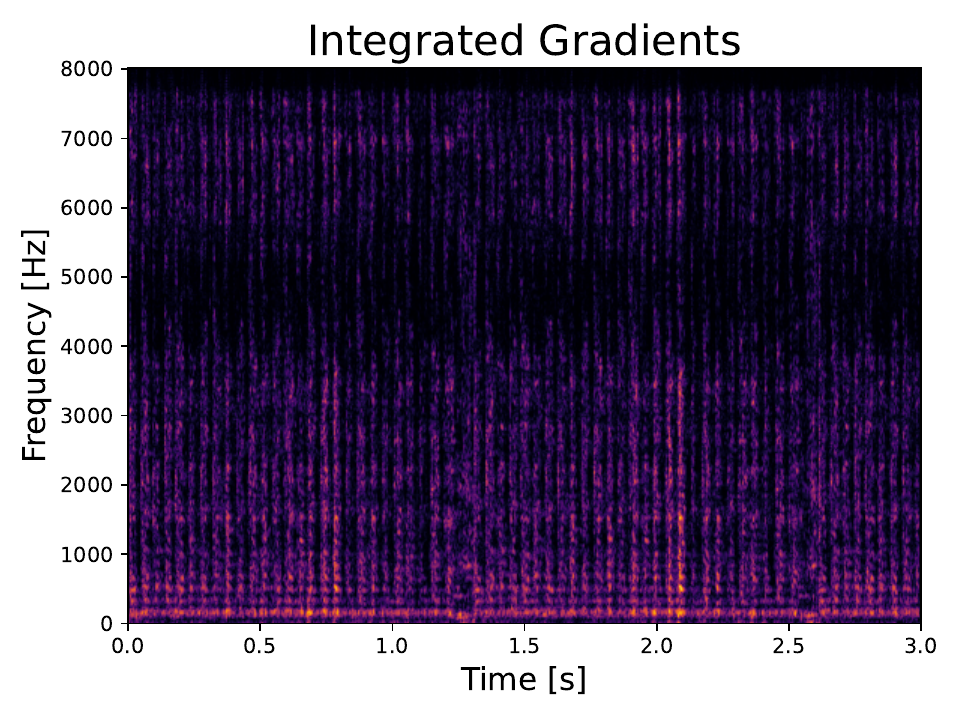}};
        \node[right of=p1, xshift=3.5cm] (p2) {\includegraphics[width=0.26\textwidth]{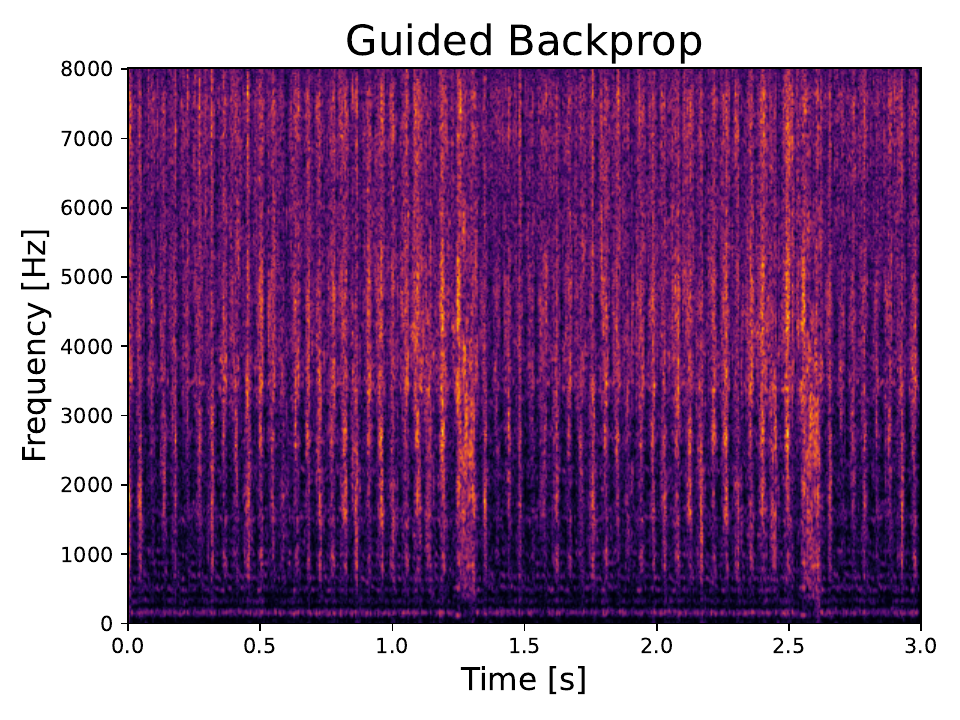}};
        \node[right of=p2, xshift=3.5cm] (p2) {\includegraphics[width=0.26\textwidth]{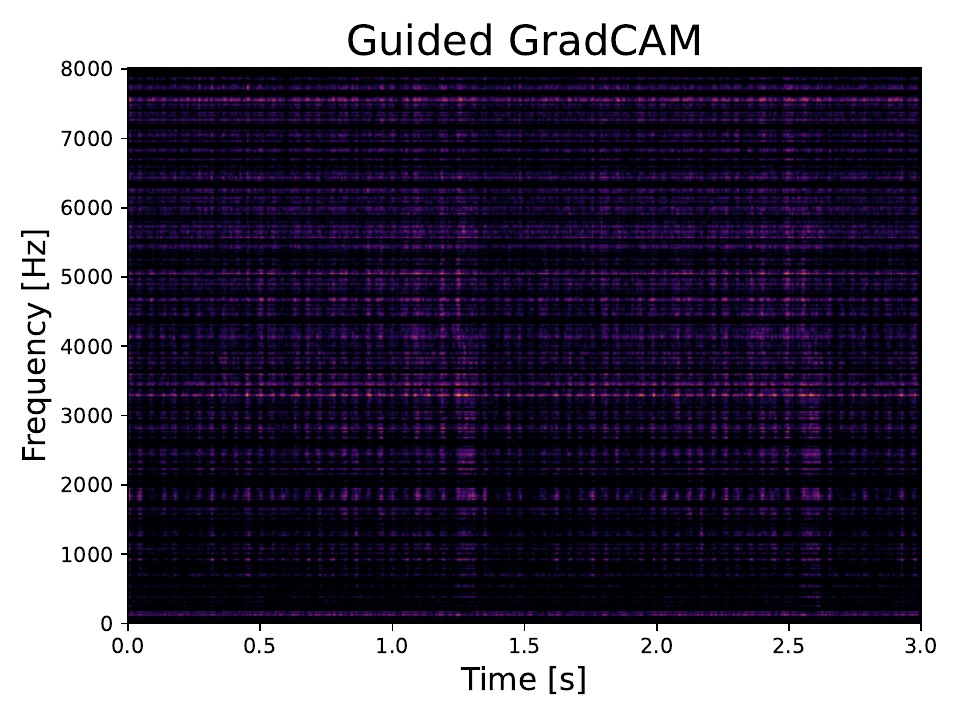}};
    \end{tikzpicture}
    }
    \vspace{-0.6cm}
    \caption{Explanations generated for a PD sample correctly classified by HuBERT. The explanations highlight different portions of the spectrogram, suggesting that understandability is difficult to achieve in this setting. From left to right: original sample, Integrated Gradients, Guided Backprop, and Guided GradCAM.}
    \label{fig:qual}
    \vspace{-0.5cm}
\end{figure*}

In contrast, there has been increasing interest in interpretability for audio and speech applications in recent years. Key works have introduced methods like layer-wise relevance propagation \cite{becker2023audiomnist}, masked additive white noise \cite{trinh18_interspeech, kavaki20_interspeech}, and Guided Backpropagation \cite{muckenhirn19_interspeech} to understand important spectrogram features. Additionally, SLIME \cite{Mishra2017LocalIM, mishra2020reliable} and AudioLIME \cite{haunschmid2020audiolime, chowdhury2021tracing} have explored feature importance within predefined regions of the spectrogram. Recent advancements, such as Listen-to-Interpret (L2I) \cite{l2i}, L-MAC \cite{paissan24lmac}, and LMAC-TD \cite{mancini2024lmactdproducingtimedomain}, have focused on generating listenable explanations in spectrogram and time domains, underscoring the value of interpretability in audio analysis.

In response to the limitations highlighted in PD detection, our work evaluates the effectiveness of XAI methods in the aforementioned context in a systematic way, particularly of perturbation and saliency-based approaches. By comparing various techniques, we aim to provide insights into their effectiveness and limitations, highlighting the potential for explainable AI in clinical applications for Parkinson's disease.

% In the audio domain, notable works on interpretability include \cite{becker2023audiomnist}, which proposed  layer-wise relevance propagation to generate saliency maps over spectrograms. Another noteworthy works include \cite{trinh18_interspeech, kavaki20_interspeech} where authors learn to identify important parts of the input spectrogram by masking additive white noise, within the context of automatic speech recognition. Additionally, \cite{won2019musictagging} proposed creating visualizations using attention layer outputs, while \cite{muckenhirn19_interspeech} suggested using Guided Backpropagation for spectrogram saliency maps.
% SLIME \cite{Mishra2017LocalIM, mishra2020reliable} proposes to divide the spectrogram into predefined time/frequency regions (akin to the superpixels in LIME \cite{ribeiro2016why} for images) and determines the feature importance for each region. AudioLIME \cite{haunschmid2020audiolime, chowdhury2021tracing}, on the other hand, defines the LIME superpixels as sources extracted from the input audio and determines a saliency score for each source. More recently \cite{l2i} proposed Listen-to-Interpret (L2I) to learn the classifier relevance for Non-Negative Matrix factorization (NMF) \cite{Lee1999LearningTP} dictionaries, via a decoder trained to estimate NMF activations. This work is particularly relevant to our method, as we also train a decoder. Consequently, we include detailed comparisons with L2I in our experiments.
\vspace{-.02cm}

\section{Methodology}
Our methodology to compare and evaluate XAI methods consists of (i) obtaining attributions and saliency maps using mainstream interpretability techniques (Section \ref{sec:m1}), (ii) evaluating the explanatory power of such saliency maps (and their combination) quantitatively via a range of metrics defined in the literature (Sections \ref{sec:m2} and \ref{sec:m3}), and (iii) evaluate the information conveyed by saliency maps for the PD detection task from an auxiliary classifier (Section \ref{sec:m4}).
%This section describes the step-by-step procedure used to investigate the interpretability of saliency maps for Parkinson’s disease (PD) detection from speech. 
\vspace{-.1cm} % we should be ok already
\subsection{Transforming Waveforms for Gradient-Based Attributions} \label{sec:m1}

Many SSL-based pre-trained models for audio operate directly on the raw waveform. This is the case also for our PD detection model, HuBERT \cite{hubert}. In this case, explaining the model predictions in the classifier's input domain (i.e. on the waveform $X_w$) results in explanations that are hard to interpret visually. We transform the waveforms into a time-frequency representation $X_f$ using the short-time Fourier transform (STFT).
% Given that our model operates directly on raw waveform inputs, we begin by transforming the audio data into short-time Fourier transform (STFT) representations with gradients enabled. 
This transformation allows us to compute \emph{attributions} ($A$) on the spectrogram of the input audio via saliency-based interpretability techniques. Before inputting the audio to the model, we convert the spectrogram back to the time domain using the inverse short-time Fourier transform (ISTFT) and the phase information of the original sample $X_w$. We apply the following interpretability techniques to generate these maps: Saliency \cite{saliency}, SmoothGrad \cite{smoothgrad}, Integrated Gradients (IG) \cite{ig}, Guided GradCAM \cite{gradcam}, Guided Backpropagation \cite{guidedbackprop}, and Guided SHAP \cite{shap}.
\begin{table*}[t]
\caption{PD quantitative evaluation results of HuBERT Base model averaged over 10-folds on s-PC-GITA.}
\label{tab:hubert_interpretability}
\centering
\begin{tabular}{l|ccccccc}
\toprule
\textbf{Metric} & AI ($\uparrow$) & AD ($\downarrow$) & AG ($\uparrow$) & FF ($\uparrow$) & Fid-In ($\uparrow$) & SPS ($\uparrow$) & COMP ($\downarrow$) \\
\midrule
Saliency & 74.66 $\pm$ 7.06 & 1.80 $\pm$ 2.22 & 64.09 $\pm$ 13.01 & 0.004 $\pm$ 0.003 & 82.82 $\pm$ 14.27 & 0.69 $\pm$ 0.02 & 11.98 $\pm$ 0.09 \\
Smoothgrad & 75.10 $\pm$ 8.43 & 1.85 $\pm$ 1.86 & 55.54 $\pm$ 13.42 & 0.004 $\pm$ 0.002 & 81.89 $\pm$ 11.76 & 0.50 $\pm$ 0.02 & 12.51 $\pm$ 0.05 \\
Guided GradCAM & 64.99 $\pm$ 11.71 & 5.24 $\pm$ 6.74 & 55.85 $\pm$ 16.06 & 0.001 $\pm$ 0.001 & \textbf{82.85 $\pm$ 13.83} & \textbf{0.83 $\pm$ 0.02} & \textbf{10.69 $\pm$ 0.12} \\
Guided Backprop & 75.21 $\pm$ 6.71 & 1.62 $\pm$ 2.06 & 64.52 $\pm$ 12.90 & 0.005 $\pm$ 0.004 & 82.67 $\pm$ 14.03 & 0.70 $\pm$ 0.02 & 11.94 $\pm$ 0.11 \\
Integrated Gradients & \textbf{78.83 $\pm$ 6.56} & \textbf{1.22 $\pm$ 1.72} & \textbf{69.35 $\pm$ 9.20} & \textbf{0.013 $\pm$ 0.008} & 81.93 $\pm$ 14.32 & 0.77 $\pm$ 0.01 & 11.69 $\pm$ 0.0 \\
Gradient SHAP & 76.55 $\pm$ 7.59 & 2.26 $\pm$ 3.65 & 67.46 $\pm$ 10.60 & 0.004 $\pm$ 0.003 & 81.93 $\pm$ 13.75 & 0.69 $\pm$ 0.01 & 11.99 $\pm$ 0.06 \\
\bottomrule
\end{tabular}
\vspace{-0.5cm}
\end{table*}

\subsection{Quantitative Analysis through Explainability Metrics} \label{sec:m2}

To quantitatively evaluate the quality of the saliency maps, we employ several explainability metrics. Specifically, we adopt metrics previously used in the L-MAC~\cite{paissan24lmac}, LMAC-ZS~\cite{paissan2024listenablemapszeroshotaudio} and  LMAC-TD~\cite{mancini2024lmactdproducingtimedomain} studies.  We use Average Increase (AI), which measures the percentage of samples for which we observe an increase in the classifier’s confidence for the interpretation with respect to the input sample, and Average Decrease (AD), which measures the confidence drop when masking the input with the mask, and Average Gain (AG), similar to AI. Beyond these metrics, we use the Faithfulness (FF) metric defined in the L2I paper \cite{l2i} and the input fidelity (Fid-In) metric defined in the PIQ paper \cite{paissan2023posthoc}. We also use the Sparseness (SPS) \cite{chalasani2020concise} and Complexity (COMP) \cite{bhatt2020evaluating} metrics to evaluate the conciseness of the explanations. 
We invite the reader to refer to the cited papers for further details. % \EM{Copied from ICASSP paper, should we modify it a bit?}
\vspace{-.2cm}
\subsection{Exploring Overlap in Explainability Methods} \label{sec:m3}

A reasonable expectation when explaining a classifier is that good explanations are aligned with the classifier's predictions, regardless of the process that generated them.  To verify this hypothesis, we investigate the overlap among different attribution methods to assess whether combining them provides further insights into the model's decisive process.
Given a dataset element $(X_f, y)$, we extract two saliency maps $A_1$, $A_2$ using two different explanation techniques (e.g. Saliency and Smoothgrad) and combine them via intersection and union.
%The intuition here is that each explainability method may highlight different important portions of the input. We measure overlap by computing the normalized intersection between two attribution maps as follows:
% % \begin{align}
% %     OA = \frac{\text{A1} \times \text{A2}}{\text{A1} + \text{A2} + 1 \times 10^{-10}}
% % \end{align}
% where OA stands for \textit{overlap attribution}, \textit{A1} and \textit{A2} for attributions compute through two different methods. 
An increase in explainability metrics through overlapping attributions would suggest that combining methods captures more comprehensive feature representations.% \EM{Add Figure for which, for the same sample we have: original spec, mask1, mask2, saliency1, saliency2 to justify overlap exploration}
\vspace{-.20cm}
\subsection{Classification Using Saliency Maps and Selective Metrics} \label{sec:m4}

Finally, to examine whether the saliency maps derived from each attribution method highlight information pertinent to distinguishing between PD and healthy controls (HC), we train a classifier on the generated explanations. %This experiment tests whether the generated saliency maps contain discriminative features for PD classification.
This test is similar to RemOve And Retrain \cite{roar}; in our case, however, we measure the amount of information in the explanations in a single iteration.  % independently of their interpretability for human understanding.
Additionally, we introduce a new set of metrics, called \emph{Selective Metrics}, that scales the standard classification metrics depending on the explanation selectiveness. These metrics penalize explanations replicating the input audio without selectively highlighting the portions of the input relevant to the classification. % The best explanation techniques with respect to the selective metrics are the ones that achieve the maximum balance between classification metrics and selectiveness.
% Standard metrics alone may not reflect this selectivity, as high performance could indicate that the classifier is simply using all available information, rather than focusing on the relevant portions highlighted by the saliency maps. To address this, 
% we define \textbf{Selective Metrics}, which evaluate the classifier's ability to achieve high performance while relying on minimal, targeted information.

To define selective metrics, we combine the classification performance (e.g. accuracy) and the average of the attribution mask. Numerically, for a dataset $\mathcal{D}=\{(X_f, y)_i\}_{i=1}^N$ of $N$ elements and metric $M(X_f, y)$ (i.e. accuracy, where $X_f$ is the STFT domain input audio), this computes as:
\begin{equation}
    S_\text{M}(\mathcal{D}) = \frac{1}{N}\sum_{(X_f, y) \in \mathcal{D}} M(X_f, y) \left(1 - \text{Av}(A) \right)
\end{equation}
where $\text{Av}(\cdot)$ denotes the mean that is calculated on the normalized attributions $A$ (which is in $[0, 1]$).
% Selective metrics are a variation of standard metrics in which we multiply each metric by a function of the mask energy. Specifically, $(1 - MaskMean)$, where $MaskMean$ represents the mean value of the saliency map. 
This adjustment rewards classifiers that achieve high accuracy while focusing on smaller, relevant input parts. This is particularly relevant to facilitate a comparison between attribution strategies. %\EM{Fra: Add details }% , defined as:

% \begin{align}
%     Selective Accuracy = Accuracy \times (1-MaskMean)
% \end{align}
% \begin{align}
%     Selective F1Score = F1Score \times (1-MaskMean)
% \end{align}
% These selective metrics help determine whether the classifier effectively identifies and uses only the key features indicated by the saliency maps.

\section{Experiments}
\subsection{Dataset}
Standard PC-GITA (s-PC-GITA) is a dataset consisting of recordings from 100 individuals, split evenly between two groups: 50 people with Parkinson’s disease (PD) and 50 healthy controls (HC). Each group includes 25 men and 25 women, with the PD group diagnosed by a neurologist, while the HC group shows no signs of PD or other neurodegenerative conditions. Participants range in age from 31 to 86 years, with recordings captured in a sound-proof booth at Clínica Noel in Medellín, Colombia. The original recordings were sampled at 44.1 kHz with 16-bit resolution and downsampled to 16 kHz for this study, as in \cite{narendra2021detection} and \cite{laquatra24_interspeech}. We use the same splits as those employed by \cite{laquatra24_interspeech} for inference on the test sets of the 10 folds and present the results averaged across these folds; our replicated results, which align with the original paper, are shown in Table \ref{tab:pd_results}. As in \cite{laquatra24_interspeech}, the speech tasks from both datasets considered in this work are diadochokinetic (DDK) exercises, read sentences, and monologues. Additional dataset details are available in \cite{orozco2014new}.

\begin{table}[h!]
    \centering
    \vspace{-0.3cm}
    \caption{PD results averaged over 10-folds on s-PC-GITA. Mean value and standard deviation are reported.}
    \begin{tabular}{l|cc}
        \toprule
        \textbf{Model} & Accuracy & F1-score \\ 
        \midrule
        HuBERT Base~\cite{hubert} & 81.32 $\pm$ 8.06 & 81.03 $\pm$ 8.33 \\ 
        WavLM Base~\cite{wavlm} & 82.10 $\pm$ 7.94 & 81.90 $\pm$ 8.09 \\ 
        \bottomrule
    \end{tabular}
    \label{tab:pd_results}
    \vspace{-0.3cm}
\end{table}

\subsection{PD Detector}

\begin{table*}[h!]
\caption{Results of PD detection using saliency maps.}
\label{tab:performance_metrics}
\centering
\resizebox{0.8\textwidth}{!}{
\begin{tabular}{l|cccccc}
\toprule
\textbf{Metric} & Accuracy  & Selective Accuracy  & F1-score & Selective F1-score  & Mask Mean  & Mask Std  \\
\midrule
Saliency & 0.87 $\pm$ 0.12 & 0.86 $\pm$ 0.10 & 0.87 $\pm$ 0.14 & 0.85 $\pm$ 0.13 & 0.017 & 0.037 \\
Smoothgrad & 0.86 $\pm$ 0.11 & 0.85 $\pm$ 0.11 & 0.86 $\pm$ 0.13 & 0.85 $\pm$ 0.12 & 0.013 & 0.020 \\
Guided GradCAM & 0.78 $\pm$ 0.09 & 0.78 $\pm$ 0.09  & 0.78 $\pm$ 0.10 & 0.78 $\pm$ 0.10 & 0.002 & 0.010 \\
Guided Backprop & 0.87 $\pm$ 0.10 & 0.86 $\pm$ 0.10 & 0.85 $\pm$ 0.14 & 0.84 $\pm$  0.13 & 0.016 & 0.035 \\
Integrated Gradients & \textbf{0.89} $\pm$ \textbf{0.08} & \textbf{0.88} $\pm$ \textbf{0.08} & \textbf{0.89} $\pm$ \textbf{0.09} & \textbf{0.88} $\pm$ \textbf{0.09} & 0.008 & 0.023 \\
Gradient SHAP & 0.84 $\pm$  0.13 & 0.83 $\pm$  0.13 & 0.85 $\pm$ 0.13 & 0.84 $\pm$ 0.13 & 0.007 & 0.019 \\
\midrule
Original Spectrograms & 0.82 $\pm$ 
 0.12 & - & 0.85 $\pm$ 0.10 & - & - & -  \\
\bottomrule
\end{tabular}
}
\vspace{-0.5cm}
\end{table*}

In this study, we build upon recent advancements in the field by relying on a study that explores exploiting foundation models and speech enhancement for Parkinson’s disease detection from speech in real-world operative conditions \cite{laquatra24_interspeech}. In \cite{laquatra24_interspeech}, they use SSL models HuBERT\cite{hubert} and WavLM\cite{wavlm} as foundational backbones for PD detection, extracting high-level representations from raw waveform inputs through a convolutional and transformer-based encoder. By leveraging pre-trained layers optimized through dynamic weighted summation and an attention pooling head, the model captures discriminative features essential for PD detection, refining these representations with fully connected layers to enhance task-specific accuracy. Following this work, we replicate their results using only the s-PC-GITA datasets and apply explainability techniques to the model.

\subsection{Saliency Maps Classifier}
We employ a CNN14 classifier \cite{kong2020panns} to evaluate saliency maps, taking log-mel spectrograms as input. 
% We compute the log-magnitude spectrograms of the saliency maps during the computation of the attributions and apply necessary transformations before feeding them to the CNN14 model. 
This classifier is trained for a binary classification task with the following hyperparameters: batch size of 32, learning rate of 0.002, and 50 training epochs.

To train and evaluate the classifier, we construct a dataset from the masks computed on the original test set. For each fold, this dataset is split into training, validation, and test sets with a ratio of 70/15/15, ensuring that the label distribution is stratified to maintain balanced classes across the splits.
For comparison, we also train the CNN14 classifier on the original spectrograms from the test set to evaluate performance differences between saliency map-based inputs and the unmodified spectrograms. The CNN14 classifier training is done using the SpeechBrain 1.0 toolkit \cite{ravanelli2024open}.

\section{Results}
As shown in Table \ref{tab:pd_results}, both the HuBERT Base and WavLM Base models obtain comparable results on PD detection. For this reason, we conduct our experiments on HuBERT in this paper. We note that the same experimental protocol and methodology can be applied to interpreting WavLM.
Our code is publicly available and can be accessed through our companion website\footnote{\url{https://helemanc.github.io/parkinsons-speech-xai/}}, together with additional spectrogram visualizations.
%Therefore, we report the results obtained using the HuBERT model here, while results for WavLM and additional visualizations can be found on our companion website\footnote{\EM{Add link}}. Our code is publicly available\footnote{\href{https://github.com/helemanc/parkinsons-speech-interpretability}{https://github.com/helemanc/parkinsons-speech-xai}}.

\vspace{-0.2cm}
\subsection{Faithfulness Metrics}
In Table \ref{tab:hubert_interpretability}, we present the quantitative evaluation results of the HuBERT-base model averaged over 10 folds on s-PC-GITA. We observe that all the XAI approaches show comparable performance overall. The variability in the metrics can be attributed to the high sensitivity of classifier representations to individual subjects, as evidenced by the high standard deviations in Table \ref{tab:pd_results}. This subject-specific variability leads to fluctuations in faithfulness metrics across folds comparable to those observed on the classification performance. For AI, AD, AG, and FF, Integrated Gradients performs best, while for Fid-In, SPS, and COMP, Guided Grad-CAM stands out. This also aligns with the results presented in Table \ref{tab:performance_metrics}, in which Guided Grad-CAM shows the lowest mask mean, aligning with its superior performance in SPS and COMP. We note that  the small FF values is due to classifier's uncertainty, since we observed small values in the predicted logit outputs. Overall, we see that the explanations are aligned with the classifier, suggesting that masking the audio spectrogram can be an effective way of highlighting the regions of the input audio associated with the predicted label.
\vspace{-.1cm}
\subsection{Overlap between Explainability Methods}
\begin{figure}
    \centering
    \includegraphics[width=0.75\linewidth]{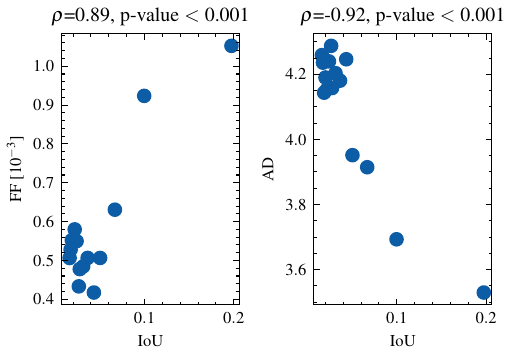}
    \vspace{-0.4cm}
    \caption{Analysis of the correlation between explanations overlap among interpretability techniques and faithfulness metrics. Combining interpretability techniques is most effective when there is already significant overlap between attribution masks.}
    \label{fig:iouvsff}
    \vspace{-0.6cm}
\end{figure}

Fig. \ref{fig:iouvsff} shows a scatter plot illustrating how faithfulness metrics vary with increasing intersection-over-union (IoU) between explanations from different methods, specifically using intersection as the combination strategy. For the results of the overlap based on the union strategy, we do not observe a linear trend with increasing IoU values. However, we observe that for some metrics (e.g. AD), union is a more effective strategy on average. We focused on two metrics, AD and FF, which showed the most variation across methods. A similar trend is observed on other metrics; we invite the reader to refer to the companion website for the results. Overall, the trend suggests that combining attribution strategies is most effective when the attributions are already well-aligned, supporting our hypothesis that greater mask overlap improves faithfulness metrics.
% Fig. \ref{fig:iouvsff} presents a scatter plot illustrating how the faithfulness metrics change with increasing intersection-over-union (IoU) between explanations from different methods. The scatter plot refers to the use of intersection as a combination strategy. We do not report the results obtained via union, since the trend observe in here is different from what we observe with intersection  We focused on two metrics, AD and FF, which showed the most variation across methods. A similar trend is observed on other metrics; we invite the reader to refer to the companion website for the results. The general trend reveals that combining multiple attribution strategies is most effective when the attributions are mostly aligned in the first place. This suggests that a greater overlap among the masks leads to better faithfulness metrics, confirming our hypothesis. \EM{Change comment as soon as we have also the results with SUM}

\subsection{Saliency Maps Classifier}
In Table \ref{tab:performance_metrics}, we report the classification accuracy and F1-score, together with the corresponding selective metrics. In most cases, training the classifier on the explanations results in higher classification performance with respect to training on the original data. This suggests that training on the explanations provides better generalization capabilities, possibly because by incorporating explanations, the classifier gains reduced context dependency, thus enhancing its generalization capabilities. Among the XAI methods, Integrated Gradients stands out, excelling in faithfulness metrics and achieving the highest selective accuracy, confirming its overall effectiveness compared to other strategies. % Additionally, we observe a positive correlation between selective metrics and FF, suggesting that the most faithful XAI techniques retain most of the information relevant for the task.

\subsection{Qualitative Analysis}
Quantitative results show that explanations align with the classifier outputs and provide helpful information for distinguishing between PD and HC. In Figure \ref{fig:qual}, we present a sample explanation generated with Guided GradCAM, Integrated Gradients and their combination. In general, we observed that the saliency maps focus on high-frequency regions, potentially reflecting attention to specific phonemes. Nonetheless, these maps are not easily interpretable by humans. We conclude that despite current XAI techniques provide faithful explanations as spectrograms, further research is needed to render explanations more insightful for domain experts. Relevant works in this direction include \cite{gupta24b_interspeech}, which relies on additional data modalities.

\section{Conclusion}
In this paper, we show that popular post-hoc explanation methods can generate faithful explanations for PD detection. Nonetheless, they fail to generate explanations that domain experts can easily understand. We, therefore, suggest that future work should explore approaches that would simplify an input spectrogram such as semantically enriching the spectrogram through phoneme discretization of the spectrogram, creating a direct link between speech biomarkers in PD research and the saliency maps, or investigating methods like listenable explanations (similar to what is proposed in L-MAC) for more intuitive insights.

\vspace{-0.1cm}
\section*{Acknowledgment}

This work was partially supported by project
``FAIR - Future Artificial Intelligence Research'' -- Spoke 8 ``Pervasive AI’’, under the European Commission's NextGeneration EU programme, PNRR -- M4C2 -- Investimento 1.3, Partenariato Esteso (PE00000013).

%\section*{References}
\bibliographystyle{IEEEtran}
\bibliography{refs}

\begin{thebibliography}{10}
\providecommand{\url}[1]{#1}
\csname url@samestyle\endcsname
\providecommand{\newblock}{\relax}
\providecommand{\bibinfo}[2]{#2}
\providecommand{\BIBentrySTDinterwordspacing}{\spaceskip=0pt\relax}
\providecommand{\BIBentryALTinterwordstretchfactor}{4}
\providecommand{\BIBentryALTinterwordspacing}{\spaceskip=\fontdimen2\font plus
\BIBentryALTinterwordstretchfactor\fontdimen3\font minus \fontdimen4\font\relax}
\providecommand{\BIBforeignlanguage}[2]{{%
\expandafter\ifx\csname l@#1\endcsname\relax
\typeout{** WARNING: IEEEtran.bst: No hyphenation pattern has been}%
\typeout{** loaded for the language `#1'. Using the pattern for}%
\typeout{** the default language instead.}%
\else
\language=\csname l@#1\endcsname
\fi
#2}}
\providecommand{\BIBdecl}{\relax}
\BIBdecl

\bibitem{hornykiewicz1998biochemical}
O.~Hornykiewicz, ``Biochemical aspects of parkinson's disease,'' \emph{Neurology}, no. 2\_suppl\_2, pp. S2--S9, 1998.

\bibitem{poewe2017parkinson}
W.~Poewe, K.~Seppi, C.~M. Tanner, G.~M. Halliday, P.~Brundin, J.~Volkmann, A.-E. Schrag, and A.~E. Lang, ``Parkinson disease,'' \emph{Nature reviews Disease primers}, no.~1, pp. 1--21, 2017.

\bibitem{de2000prevalence}
M.~d. De~Rijk, L.~Launer, K.~Berger, M.~Breteler, J.~Dartigues, M.~Baldereschi, L.~Fratiglioni, A.~Lobo, J.~Martinez-Lage, C.~Trenkwalder \emph{et~al.}, ``Prevalence of parkinson's disease in europe: A collaborative study of population-based cohorts. neurologic diseases in the elderly research group.'' \emph{Neurology}, no. 11 Suppl 5, pp. S21--3, 2000.

\bibitem{pinto2004treatments}
S.~Pinto, C.~Ozsancak, E.~Tripoliti, S.~Thobois, P.~Limousin-Dowsey, and P.~Auzou, ``Treatments for dysarthria in parkinson's disease,'' \emph{The Lancet Neurology}, no.~9, pp. 547--556, 2004.

\bibitem{rusz2013imprecise}
J.~Rusz, R.~Cmejla, T.~Tykalova, H.~Ruzickova, J.~Klempir, V.~Majerova, J.~Picmausova, J.~Roth, and E.~Ruzicka, ``Imprecise vowel articulation as a potential early marker of parkinson's disease: effect of speaking task,'' \emph{The Journal of the Acoustical Society of America}, no.~3, pp. 2171--2181, 2013.

\bibitem{laquatra24_interspeech}
M.~{La Quatra}, M.~F. Turco, T.~Svendsen, G.~Salvi, J.~R. Orozco-Arroyave, and S.~M. Siniscalchi, ``Exploiting foundation models and speech enhancement for parkinson's disease detection from speech in real-world operative conditions,'' in \emph{Interspeech 2024}, 2024.

\bibitem{molnar2022}
C.~Molnar, \emph{Interpretable Machine Learning}, 2nd~ed., 2022.

\bibitem{MERSHA2024128111}
M.~Mersha, K.~Lam, J.~Wood, A.~K. AlShami, and J.~Kalita, ``Explainable artificial intelligence: A survey of needs, techniques, applications, and future direction,'' \emph{Neurocomputing}, 2024.

\bibitem{sonawane2021speech}
B.~Sonawane and P.~Sharma, ``Speech-based solution to parkinson’s disease management,'' \emph{Multimedia Tools and Applications}, no.~19, pp. 29\,437--29\,451, 2021.

\bibitem{rios2022end}
C.~D. Rios-Urrego, S.~A. Moreno-Acevedo, E.~N{\"o}th, and J.~R. Orozco-Arroyave, ``End-to-end parkinson’s disease detection using a deep convolutional recurrent network,'' in \emph{International Conference on Text, Speech, and Dialogue}.\hskip 1em plus 0.5em minus 0.4em\relax Springer, 2022, pp. 326--338.

\bibitem{vasquez2021transfer}
J.~C. V{\'a}squez-Correa, C.~D. Rios-Urrego, T.~Arias-Vergara, M.~Schuster, J.~Rusz, E.~Noeth, and J.~R. Orozco-Arroyave, ``Transfer learning helps to improve the accuracy to classify patients with different speech disorders in different languages,'' \emph{Pattern Recognition Letters}, pp. 272--279, 2021.

\bibitem{quan2022end}
C.~Quan, K.~Ren, Z.~Luo, Z.~Chen, and Y.~Ling, ``End-to-end deep learning approach for parkinson’s disease detection from speech signals,'' \emph{Biocybernetics and Biomedical Engineering}, no.~2, pp. 556--574, 2022.

\bibitem{FAVARO2023107559}
A.~Favaro, Y.-T. Tsai, A.~Butala, T.~Thebaud, J.~Villalba, N.~Dehak, and L.~Moro-Velázquez, ``Interpretable speech features vs. dnn embeddings: What to use in the automatic assessment of parkinson’s disease in multi-lingual scenarios,'' \emph{Computers in Biology and Medicine}, p. 107559, 2023.

\bibitem{10.3389/fneur.2023.1142642}
A.~Favaro, L.~Moro-Velázquez, A.~Butala, C.~Motley, T.~Cao, R.~D. Stevens, J.~Villalba, and N.~Dehak, ``Multilingual evaluation of interpretable biomarkers to represent language and speech patterns in parkinson's disease,'' \emph{Frontiers in Neurology}, 2023.

\bibitem{maffia2024automatic}
M.~Maffia, L.~Schettino, and V.~N. Vitale, ``Automatic detection of parkinson’s disease with connected speech acoustic features: towards a linguistically interpretable approach,'' in \emph{Proceedings of the 9th Italian Conference on Computational Linguistics CLiC-it 2023: Venice, Italy, November 30-December 2, 2023}.\hskip 1em plus 0.5em minus 0.4em\relax Accademia University Press.

\bibitem{becker2023audiomnist}
S.~Becker, J.~Vielhaben, M.~Ackermann, K.-R. Müller, S.~Lapuschkin, and W.~Samek, ``Audiomnist: Exploring explainable artificial intelligence for audio analysis on a simple benchmark,'' 2023.

\bibitem{trinh18_interspeech}
V.~A. Trinh, B.~McFee, and M.~I. Mandel, ``{Bubble Cooperative Networks for Identifying Important Speech Cues},'' in \emph{Proc. Interspeech 2018}, 2018.

\bibitem{kavaki20_interspeech}
H.~S. Kavaki and M.~I. Mandel, ``{Identifying Important Time-Frequency Locations in Continuous Speech Utterances},'' in \emph{Proc. Interspeech 2020}, 2020.

\bibitem{muckenhirn19_interspeech}
H.~Muckenhirn, V.~Abrol, M.~Magimai-Doss, and S.~Marcel, ``{Understanding and Visualizing Raw Waveform-Based CNNs},'' in \emph{Proc. Interspeech 2019}, 2019, pp. 2345--2349.

\bibitem{Mishra2017LocalIM}
S.~Mishra, B.~L. Sturm, and S.~Dixon, ``Local interpretable model-agnostic explanations for music content analysis,'' in \emph{International Society for Music Information Retrieval Conference}, 2017.

\bibitem{mishra2020reliable}
S.~Mishra, E.~Benetos, B.~L. Sturm, and S.~Dixon, ``Reliable local explanations for machine listening,'' 2020.

\bibitem{haunschmid2020audiolime}
V.~Haunschmid, E.~Manilow, and G.~Widmer, ``audiolime: Listenable explanations using source separation,'' 2020.

\bibitem{chowdhury2021tracing}
S.~Chowdhury, V.~Praher, and G.~Widmer, ``Tracing back music emotion predictions to sound sources and intuitive perceptual qualities,'' 2021.

\bibitem{l2i}
J.~Parekh, S.~Parekh, P.~Mozharovskyi, F.~d\textquotesingle Alch\'{e}-Buc, and G.~Richard, ``Listen to interpret: Post-hoc interpretability for audio networks with nmf,'' in \emph{Advances in Neural Information Processing Systems}, 2022, pp. 35\,270--35\,283.

\bibitem{paissan24lmac}
F.~Paissan, M.~Ravanelli, and C.~Subakan, ``Listenable maps for audio classifiers,'' in \emph{Proceedings of the 41st International Conference on Machine Learning}, 2024, pp. 39\,009--39\,021.

\bibitem{mancini2024lmactdproducingtimedomain}
E.~Mancini, F.~Paissan, M.~Ravanelli, and C.~Subakan, ``Lmac-td: Producing time domain explanations for audio classifiers,'' \emph{arXiv preprint arXiv:2409.08655}, 2024.

\bibitem{hubert}
W.-N. Hsu, B.~Bolte, Y.-H.~H. Tsai, K.~Lakhotia, R.~Salakhutdinov, and A.~Mohamed, ``Hubert: Self-supervised speech representation learning by masked prediction of hidden units,'' \emph{IEEE/ACM Trans. Audio, Speech and Lang. Proc.}, 2021.

\bibitem{saliency}
K.~Simonyan, A.~Vedaldi, and A.~Zisserman, ``Deep inside convolutional networks: Visualising image classification models and saliency maps,'' 2014.

\bibitem{smoothgrad}
D.~Smilkov, N.~Thorat, B.~Kim, F.~Viégas, and M.~Wattenberg, ``Smoothgrad: removing noise by adding noise,'' 2017.

\bibitem{ig}
M.~Sundararajan, A.~Taly, and Q.~Yan, ``Axiomatic attribution for deep networks,'' in \emph{Proceedings of the 34th International Conference on Machine Learning - Volume 70}, ser. ICML'17.\hskip 1em plus 0.5em minus 0.4em\relax JMLR.org, 2017.

\bibitem{gradcam}
R.~R. Selvaraju, M.~Cogswell, A.~Das, R.~Vedantam, D.~Parikh, and D.~Batra, ``Grad-cam: Visual explanations from deep networks via gradient-based localization,'' in \emph{2017 IEEE International Conference on Computer Vision (ICCV)}, 2017.

\bibitem{guidedbackprop}
J.~T. Springenberg, A.~Dosovitskiy, T.~Brox, and M.~A. Riedmiller, ``Striving for simplicity: The all convolutional net,'' in \emph{3rd International Conference on Learning Representations, {ICLR} 2015, San Diego, CA, USA, May 7-9, 2015, Workshop Track Proceedings}, Y.~Bengio and Y.~LeCun, Eds., 2015.

\bibitem{shap}
S.~M. Lundberg and S.-I. Lee, ``A unified approach to interpreting model predictions,'' in \emph{Proceedings of the 31st International Conference on Neural Information Processing Systems}, ser. NIPS'17.\hskip 1em plus 0.5em minus 0.4em\relax Red Hook, NY, USA: Curran Associates Inc., 2017.

\bibitem{paissan2024listenablemapszeroshotaudio}
F.~Paissan, L.~D. Libera, M.~Ravanelli, and C.~Subakan, ``Listenable maps for zero-shot audio classifiers,'' \emph{arXiv preprint arXiv:2405.17615}, 2024.

\bibitem{paissan2023posthoc}
F.~Paissan, C.~Subakan, and M.~Ravanelli, ``Posthoc interpretation via quantization,'' \emph{arXiv preprint arXiv:2303.12659}, 2023.

\bibitem{chalasani2020concise}
P.~Chalasani, J.~Chen, A.~R. Chowdhury, S.~Jha, and X.~Wu, ``Concise explanations of neural networks using adversarial training,'' 2020.

\bibitem{bhatt2020evaluating}
U.~Bhatt, A.~Weller, and J.~M.~F. Moura, ``Evaluating and aggregating feature-based model explanations,'' 2020.

\bibitem{roar}
S.~Hooker, D.~Erhan, P.-J. Kindermans, and B.~Kim, ``A benchmark for interpretability methods in deep neural networks,'' 2019.

\bibitem{narendra2021detection}
N.~Narendra, B.~Schuller, and P.~Alku, ``The detection of parkinson's disease from speech using voice source information,'' \emph{IEEE/ACM Transactions on Audio, Speech, and Language Processing}, pp. 1925--1936, 2021.

\bibitem{orozco2014new}
J.~R. Orozco-Arroyave, J.~D. Arias-Londo{\~n}o, J.~F. Vargas-Bonilla, M.~C. Gonzalez-R{\'a}tiva, and E.~N{\"o}th, ``New spanish speech corpus database for the analysis of people suffering from parkinson's disease.'' in \emph{Lrec}, 2014, pp. 342--347.

\bibitem{wavlm}
S.~Chen, C.~Wang, Z.~Chen, Y.~Wu, S.~Liu, Z.~Chen, J.~Li, N.~Kanda, T.~Yoshioka, X.~Xiao, J.~Wu, L.~Zhou, S.~Ren, Y.~Qian, Y.~Qian, J.~Wu, M.~Zeng, X.~Yu, and F.~Wei, ``Wavlm: Large-scale self-supervised pre-training for full stack speech processing,'' \emph{IEEE Journal of Selected Topics in Signal Processing}, 2022.

\bibitem{kong2020panns}
Q.~Kong, Y.~Cao, T.~Iqbal, Y.~Wang, W.~Wang, and M.~D. Plumbley, ``Panns: Large-scale pretrained audio neural networks for audio pattern recognition,'' 2020.

\bibitem{ravanelli2024open}
M.~Ravanelli, T.~Parcollet, A.~Moumen, S.~de~Langen, C.~Subakan, P.~Plantinga, Y.~Wang, P.~Mousavi, L.~Della~Libera, A.~Ploujnikov \emph{et~al.}, ``Open-source conversational ai with speechbrain 1.0,'' \emph{arXiv preprint arXiv:2407.00463}, 2024.

\bibitem{gupta24b_interspeech}
S.~Gupta, M.~Ravanelli, P.~Germain, and C.~Subakan, ``Phoneme discretized saliency maps for explainable detection of ai-generated voice,'' in \emph{Interspeech 2024}, 2024, pp. 3295--3299.

\end{thebibliography}

\end{document}